\documentclass[pre,twocolumn,showpacs]{revtex4-1}

\usepackage{graphicx}
\usepackage{amssymb}
\usepackage{amsmath}
\usepackage{subfigure,wrapfig}
\usepackage{color}
\usepackage{ulem}
\usepackage{cancel}
\usepackage{enumerate}
\usepackage{cases}

\begin{document}

\title{Selection of the Taylor-Saffman Bubble does not Require Surface Tension}

\author{Giovani L.~Vasconcelos}

\affiliation{Laborat\'orio de F\'{\i}sica Te\'orica e Computacional,
Departamento de F\'{\i}sica, Universidade Federal de Pernambuco,
50670-901, Recife, Brazil.}
\author{Mark Mineev-Weinstein}
\affiliation{New Mexico Consortium, Los Alamos, NM 87544, US}

\date{\today}

\begin{abstract}

A new general class of exact solutions is presented for the time
evolution of a bubble of arbitrary initial shape in a Hele-Shaw cell
when surface tension effects are neglected. These solutions are
obtained by conformal mapping the viscous flow domain to an annulus
in an auxiliary complex-plane. It is then demonstrated that the only
stable fixed point (attractor) of the non-singular bubble dynamics
corresponds precisely to the selected pattern. This thus shows that,
contrary to the established theory, bubble selection in a Hele-Shaw
cell does not require surface tension. The solutions reported here
significantly extend previous results for a simply-connected
geometry (finger) to a doubly-connected one (bubble).  We conjecture
that the same selection rule without surface tension holds for
Hele-Shaw flows of arbitrary connectivity.  We also believe that
this mechanism can be found in other, similarly described, selection
problems.

\end{abstract}

\maketitle

{\it Introduction.} It is remarkable that numerous
processes
of pattern formation, ranging from dendritic and fractal growth to
viscous fingering and bacterial colony growth, have (after some
idealization) the same compact mathematical formulation
\cite{Langer,Kadanoff}.  Subsequent development of this formulation,
called {\it Laplacian growth},
significantly widened the list of connections
by including there 1D-turbulence and generation of complex shapes
\cite{Hastings},
quantum gravity,
integrable systems, random matrices, and conformal theories
\cite{MWZ}. The problem of great importance in some processes
mentioned above (such as dendritic growth \cite{Ivantsov} and
viscous fingering in a Hele-Shaw cell \cite{ST}) was selection of
the observed pattern from continuously many solutions.

{\it Background.}  It has been widely accepted that surface tension
is necessary for selecting a single
pattern from a continuum of
solutions
in interface dynamics \cite{Pelce} after it was first conjectured by
Saffman and Taylor in 1958 for viscous fingers in a Hele-Shaw cell
\cite{ST}.  Based on their experimental observation these authors
argued that as surface tension effects becomes negligible the only
one of the set of theoretical finger shapes which can occur is that
for which the finger propagates twice as fast as the background flow
\cite{ST}. But verifying this selection scenario was not possible
until much later because of significant mathematical difficulties to
include surface tension.
After the seminal work of Kruskal and Segur \cite{KS}, these
difficulties were finally resolved, and in the series of works by
different groups \cite{PRL86} it was shown that surface tension
indeed selects the observed pattern;
see also \cite{Pelce} and \cite{KKL}.

More recently, however, it was demonstrated in \cite{1998}, by using
time-dependent exact solutions without surface tension, that
contrary to previously mentioned works, selection of the
Saffman-Taylor finger is determined entirely by the zero surface
tension dynamics.  This result calls for a revision of the role of
surface tension in pattern selection.  But even the simple question
of whether selection without surface tension is valid beyond
simply-connected domains (considered in \cite{1998}), already
presents a mathematical challenge, namely, finding time-dependent
exact solutions for multiply-connected Hele-Shaw flows.  In the
present work we solve this problem for a doubly-connected geometry
and significantly extend the results obtained in \cite{1998} by
addressing the dynamics of a bubble in a Hele-Shaw cell instead of a
finger (which is the singular limit of an infinitely long bubble).
This extension allows us to conjecture that surface tension (when
small enough) is not required for pattern selection in Laplacian
growth in domains of arbitrary connectivity.

The pattern selection problem for an inviscid bubble dragged by a
viscous flow in a narrow gap between parallel plates (Hele-Shaw
cell) was posed by Taylor and Saffman in 1959 \cite{TS} and was
later addressed experimentally \cite{PH} and theoretically
\cite{T87}.  To be specific, the problem was how to select, from the
continuum of steady state solutions obtained for zero surface
tension, the unique bubble \cite{TS} with velocity twice the
background flow velocity.  While it has long been known that the
inclusion of surface tension leads to velocity selection \cite{T87},
we demonstrate in this paper that the selection mechanism does not
require surface tension. Instead, the selection comes about because
the selected pattern is the only stable fixed point (attractor) of
the non-singular bubble dynamics {\it without surface tension}.

{\it Problem formulation and plan of the paper.} A top-view of a
Hele-Shaw channel with lateral walls at $y=\pm \pi$ in our scaled
units and with the bubble moving to the right is shown on Fig.~1a.
The fluid (oil) velocity obeys the 2D Darcy law, ${\bf v} = -\nabla
p$, where $p$ is scaled pressure. Far from the bubble the oil flows
along the $Ox$ axis with uniform velocity, $V=1$, thus $p=-x$, when
$|x| \to \infty$.  Owing to incompressibility, $\nabla \cdot {\bf v}
= 0$, $p$ is harmonic, $\nabla^2 p = 0$, in the viscous domain,
$D(t)$, where $t$ denotes time. It is thus convenient to introduce a
complex potential, $W=\Phi+i\Psi$, where $\Phi = -p$ and $\Psi$ is
the stream function. In view of the uniform far-field velocity, one
has $W \approx z$ for $|x| \to \infty$, so $\Psi =\pm\pi$ at
$y=\pm\pi$, since  the lateral walls are streamlines. Because
pressure is constant (taken to be zero) in the inviscid bubble and
continuous across the oil/bubble interface, $\Gamma(t)$, if surface
tension is neglected, then $\Phi = 0$ at $\Gamma(t)$. The fluid
domain in the $W$-plane is shown in Fig.~1b, where the vertical slit
maps to the interface $\Gamma(t)$ and the rest of the horizontal
strip, $-\pi < \Psi < \pi$, to the
domain $D(t)$, in the physical plane (Fig.~1a).  The kinematic
identity, $V_n=v_n$, stating the equality of the normal velocities
of the interface, $V_n$, and of the fluid, $v_n$, completes the
formulation of this free-boundary problem of finding $\Gamma(t)$
given $\Gamma(0)$.

\begin{figure}[b]
\begin{center}
\subfigure[\label{fig:1a}]{\includegraphics[width=0.2\textwidth]{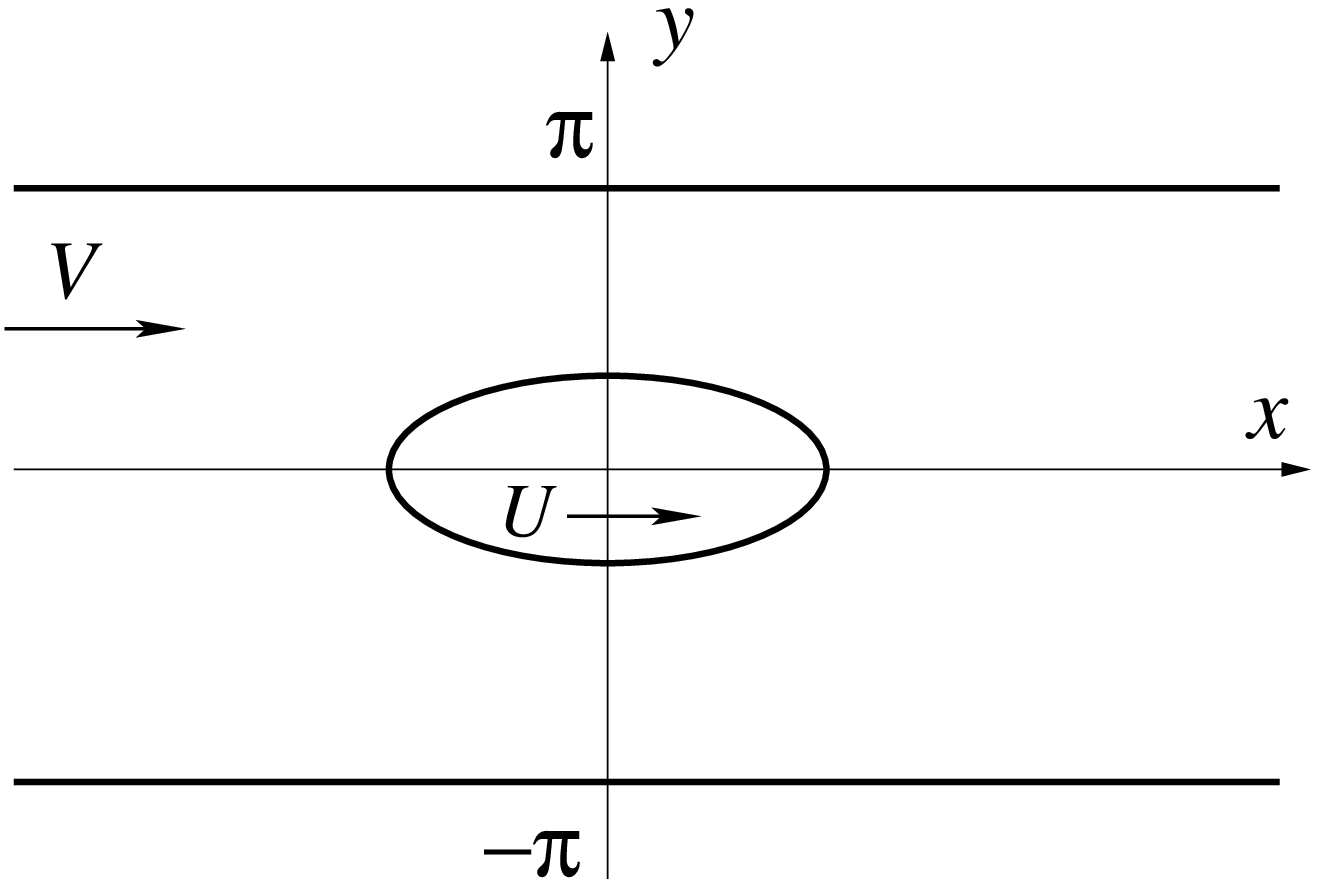}}
\subfigure[\label{fig:1b}]{\includegraphics[width=0.21\textwidth]{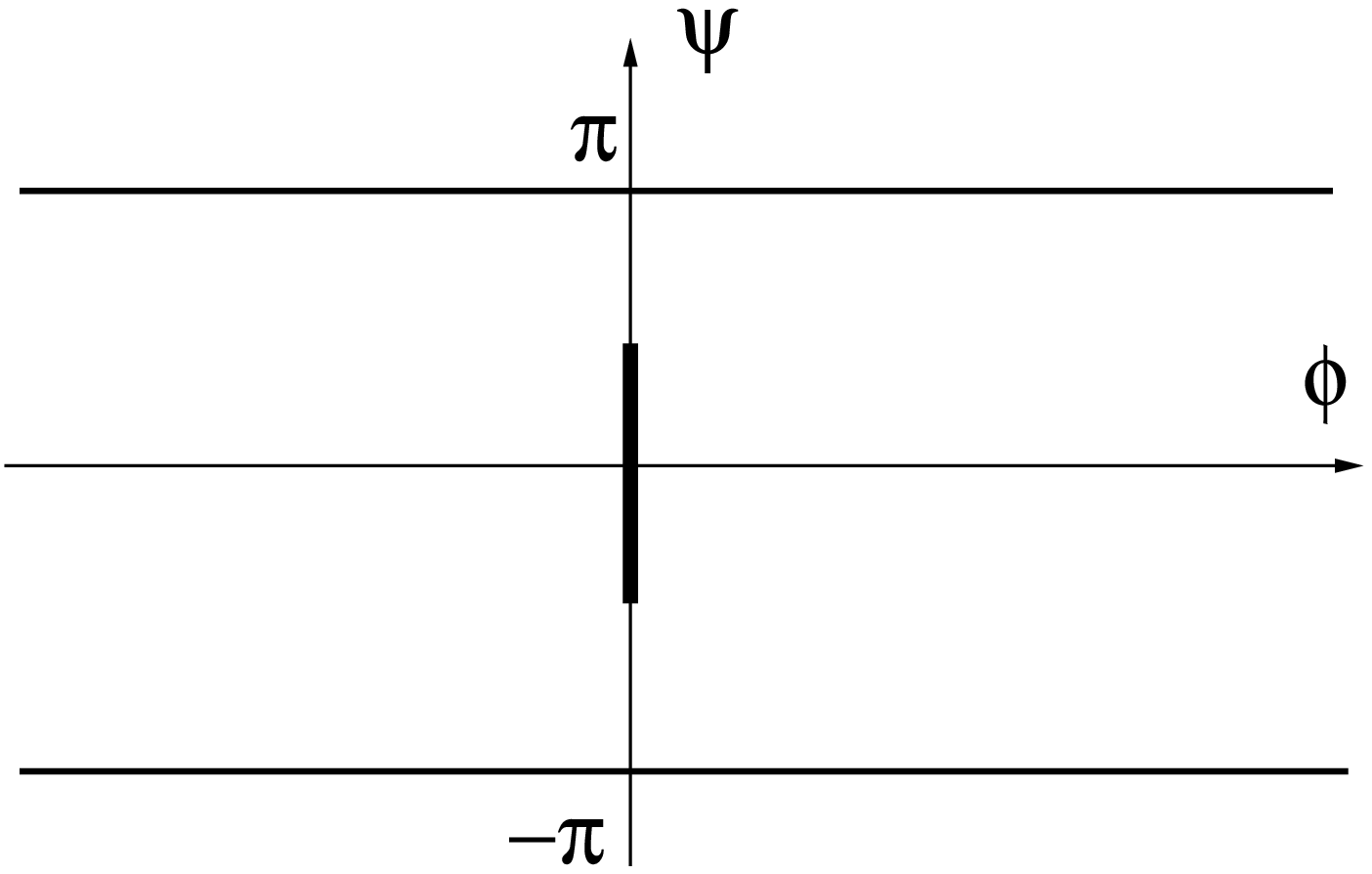}}
\subfigure[\label{fig:1c}]{\includegraphics[width=0.22\textwidth]{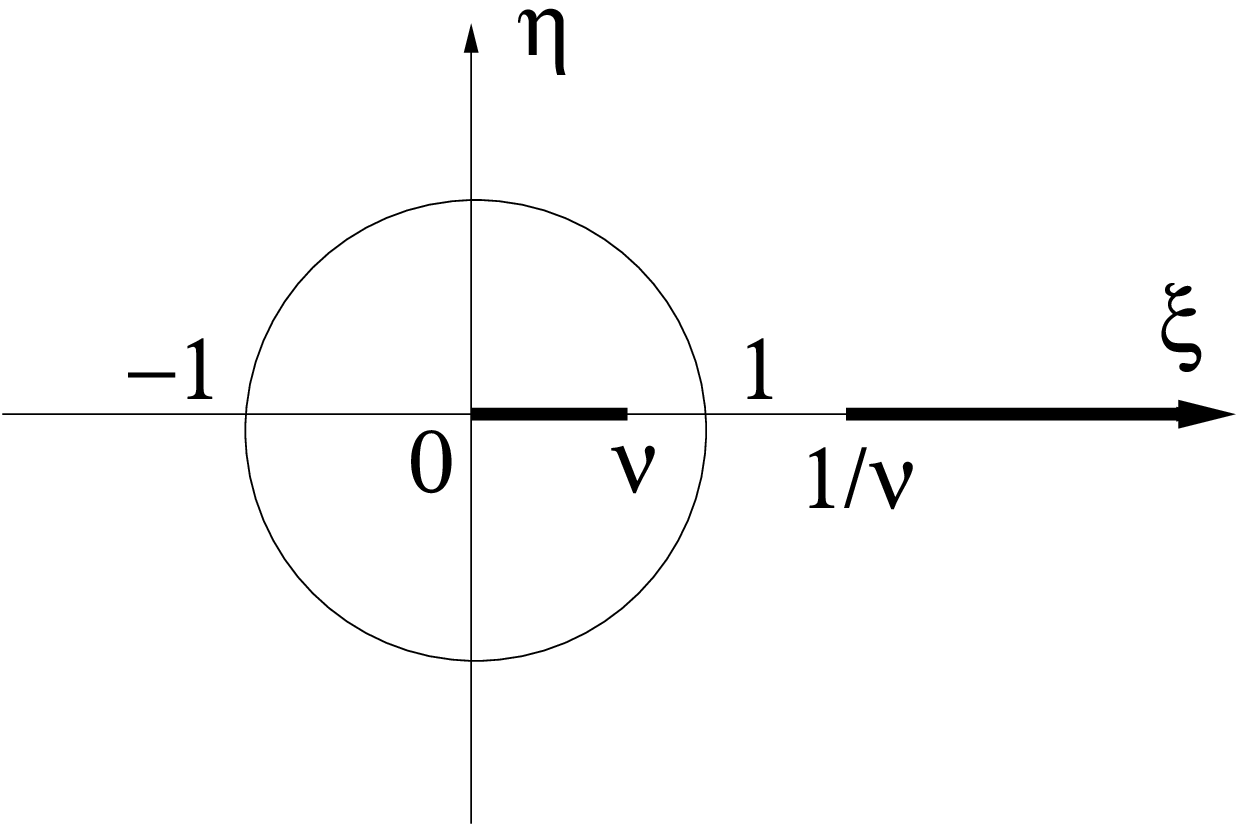}}
\end{center}
\caption{The fluid domain $D(t)$ for a moving bubble in a Hele-Shaw
channel (a), and the corresponding domains in the complex potential
plane (b)  and  in the auxiliary $\zeta$-plane (c).} \label{fig:1}
\end{figure}

This long-standing nonlinear unstable problem \cite{Gillow} was
shown to have an integrable structure and to possess deep
connections with other branches of mathematical physics \cite{MWZ}.
While numerous exact solutions were obtained for time-dependent
interfaces (listed in \cite{Gustafsson}) and for steady shapes (see
\cite{GLV2001} and references therein), almost all of them were for
simply-connected domains $D(t)$.

Here we present exact solutions for an evolving bubble in a
Hele-Shaw cell when the fluid domain  $D(t)$ is doubly connected
(see Fig.~1a). We then show that these solutions explain not only
{\it how} the moving bubble eventually reaches a stationary shape
when $t\to\infty$, but also {\it why} the  selected bubble moves
precisely twice as fast as the background flow when surface tension
effects become negligible, in agreement with \cite{PH}.

In what follows, we first address the dynamics of a bubble with a
shape symmetric with respect to the channel centerline. The exact
solutions we obtain for this simpler case illustrate our main
result, namely, that all non-singular solutions describing bubble
shapes converge to the only stable fixed point of this
infinitely-dimensional dynamical system, which is precisely the
selected member of the continuous family \cite{TS}. After that, we
extend our solutions to asymmetric bubbles and arrive at the
conclusion that $U=2$ is still the selected value.

{\it The conformal map and the equation of motion.} For a symmetric
bubble we introduce a conformal map $z = f(t,\zeta)$ from the
exterior of the unit circle with a cut in the complex plane, $\zeta
= \xi +i\eta$ (see Fig.~1c), onto the fluid domain, $D(t)$ in the
$z$-plane, $z=x+iy$.  The unit circle, $|\zeta|=1$, maps onto the
interface, $\Gamma(t)$, and the cut sides, $\zeta = \xi \pm i0$,
where $1<1/\nu(t)<\xi < \infty$, map onto the channel walls,
$y=\pm\pi$, so that $\zeta = 1/\nu$ and $\xi = +\infty$ are mapped
to $x=+\infty$  and $x = -\infty$, respectively. Thus, the polar
angle $\phi = \arg(\zeta)$ parametrizes $\Gamma(t)$: $z = f(t,
e^{i\phi})$.  It is easy to see that the complex potential,
$W=\Phi+i\Psi$, satisfying all aforementioned boundary conditions,
is
\begin{equation}
W(t,\zeta) = \log \frac{1-\nu/\zeta}{1-\nu \zeta}.
\end{equation}
Rewriting the normal velocity of the interface, $V_n$, as
\begin{equation}
V_n = V_1 l_2 - V_1 l_1 = {\cal I}m({\bar V} l) = {\cal I}m(\bar z_t
z_s),\nonumber
\end{equation}
where $l = l_1+il_2 = dz/ds$ is the unit tangent vector along the
interface, $\Gamma(t) = z(t,s)$, parametrized by an arclength, $s$,
and the subscripts stand for partial derivatives. The normal fluid
velocity we rewrite as
\begin{equation}
v_n = \partial_n \Phi = \partial_s \Psi,\nonumber
\end{equation}
by virtue of the Cauchy-Riemann condition.  Equating the two last
formulae, as required by the kinematic identity, and reparametrizing
$s \to \phi$, we obtain ${\cal I}m(\bar z_t z_\phi) = \Psi_\phi$.
Calculating $\Psi_\phi$ from (1) for $\zeta = e^{i\phi}$, we obtain
the equation for the moving interface, $z(t,\phi) = f(t,
e^{i\phi})$:
\begin{equation}
{\cal I}m(\bar z_t z_\phi) = {\cal R}e \frac{2\nu}{e^{i\phi}-\nu}.
\end{equation}
For stationary solutions, $z(t,\phi) = Ut+Z(\phi)$, where the
velocity, $U$, is a constant, equation (2) is simplified:
\begin{equation}
{\cal I}m(Z_\phi) = \frac{2\nu}{U}{\cal R}e \frac{1}{e^{i\phi}-\nu}.
\end{equation}
The solution of (3) is the sum of two logarithms:
\begin{equation}
Z = -\log(1-\nu e^{i\phi}) + \alpha \log(1-\nu e^{-i\phi}),
\end{equation}
where the coefficient at the first term equals $1$ to satisfy $W=z$
in (1) when $z \to \infty$. Expression (4) is precisely the
one-parameter family of stationary bubbles obtained in \cite{TS}.
Substituting (4) into (3), we obtain
\begin{equation}
U=\frac{2}{1+\alpha}.
\end{equation}
We will show below that all solutions with $\alpha \neq 0$ are
unstable and, if perturbed, move to the solution with $\alpha = 0$,
which corresponds to $U=2$, which is the selected value
\cite{TS,PH}.

{\it The finite-parametric solution.} Being integrable, equation (2)
possesses a rich list of exact solutions, many of which blow up in
finite time. Leaving those aside as physically non-realizable, we
present here a rich class of finite parametric {\it non-singular}
solutions (analogous of those obtained earlier \cite{1994} for
finger dynamics), which remain finite for all times:
\begin{equation}
z = \tau(t) - \log( 1-\nu(t) e^{i\phi})
+\sum_{k=0}^N\,\alpha_k\,\log(1-a_k(t) e^{-i\phi}), \label{eq:zs}
\end{equation}
where $\alpha_0 = \alpha$ and $\tau$ are both real, $a_0 =\nu$,
$|a_k|<1$ for all times, and $\alpha_k$ are constants. These
parameters must be chosen so that critical points of the conformal
map always stay inside the unit circle to prevent blow ups
\cite{footnote}. Also, the symmetry of the bubble requires that for
each term with complex $\alpha_k$ and $a_k$ in the sum there is
another term with $\bar \alpha_k$ and $\bar a_k$.  It is easy to
verify that (6) is indeed a solution of (2), where the
time-dependence of $\tau$, $\nu$, and $a_k$ is given by the
following set of $N+2$ equations:
\begin{subequations}
\begin{align}
&\beta_k = \tau + \log \bar a_k/(\bar a_k - \nu)
+\sum_{l=0}^N \alpha_l \log(1-a_l\bar a_k), \label{eq:bk}\\
&2t + 2t_0 = (1 + \alpha) \tau  -  \log (1-\nu^2)
 + \sum_{k=1}^N\,\alpha_k\,\log(1-\nu a_k), \\
&A=(|\alpha|^2-1)\log (1-\nu^2)+\sum_{k=1}^N\,\sum_{l=1}^N\,\alpha_k
\bar\alpha_l \log(1-a_k\bar a_l),
\end{align}
\end{subequations}
with $k=1,...,N$ in (\ref{eq:bk}). Here the $\beta_k$'s, the initial
time $t_0$, and the bubble area $A$ are the constants of motion.

{\it The attractor.} It follows from (7b) and (7c) that $\tau(t)$
diverges linearly with time, $\tau\to Ut$, when $t \to +\infty$.
Since $\beta_k$ is a constant, a real part of at least one logarithm
in (7a) should go to $-\infty$ in large times so as to compensate a
divergent positive $\tau(t)$.  This is possible only if all $a_k(t)
\to 0$ for $k > 0$, as $t\to \infty$.  Thus, we conclude that the
origin, $\zeta = 0$, attracts all $a_k(t)$
for $k\ge 1$. Thus, for $t\to\infty$ the only non-vanishing
parameter among the $a_k$'s is $a_0$, which we have identified with
$\nu$, meaning that at the initial time we set $a_0(0)=\nu(0)$ and
so $a_0(t)=\nu(t)$ for all times.  But in this case, the solution
(6) asymptotically approaches the family of stationary bubbles (4)
discussed above, namely
\begin{equation}
z = \frac{2t}{1+\alpha} - \log(1-\nu \, e^{i\phi}) +
\alpha\log(1-\nu e^{-i\phi}).\nonumber
\end{equation}

To test the stability of the trajectory $a_0(t)=\nu(t)$, one has to
deviate $a_0\rm (0)$ slightly from $\nu\rm (0)$.
One can readily check that the ensuing dynamics for $a_0$, $\nu$ and
$\tau$ is given in this case by
\begin{eqnarray}
&\beta_0 = \tau + \log\bar a_0/(\bar a_0 - \nu)
+\alpha\log[(1-|a_0|^2)(1-\bar a_0^2)], \nonumber \\
&2(t+t_0) = \tau - \log(1-\nu^2) + \alpha\log|1-a_0\nu|, \nonumber \\
&A = -\log(1-\nu^2) + (\alpha^2/2) \log[(1-|a_0|^2)|1-a_0^2|].
\nonumber
\end{eqnarray}
These equations clearly show that $a_0 \to 0$, $\tau \to 2t$, and
$\nu \to \sqrt{1-e^{-A}}$ when $t \to \infty$, so the interface
becomes
\begin{equation}
z=2t - \log(1-\nu e^{i\phi}).
\end{equation}
This is precisely the selected pattern with $U=2$. Thus, $\zeta=\nu$
repels nearly located singularities, which move toward $0$.
Therefore, the selected bubble, described by (8), represents the
only attractor, $\zeta = 0$, of the non-singular subset of the
finite-dimensional dynamical systems (6).

{\it The infinite-parametric solution.} Let us extend (6) to the
case of an infinite number of parameters:
\begin{equation}
z = \tau(t)+ \log\frac{1}{1-\nu(t) e^{i\phi}}
+\sum_{k=0}^N\,\int_0^{a_k(t)}\frac{\rho_k(t,w)}{w-e^{i\phi}}\,dw,
\end{equation}
where $a_0=\nu$ and $|a_k|<1$ for all $k$. Solution (9) coincides
with (6), if the functions $\rho_k(t,w)$ are constants, and contains
all previously known solutions for simply-connected and symmetric
doubly-connected cases. (Apparently one can approximate any
non-singular solution of (2) by (9), although we have not proved
this.) The constants of motion for (9), analogous to $\beta_k$ in
(7), are $B_k = f(1/\bar a_k(t))$, which yield \cite{15}
\begin{equation}
B_k = \tau(t) + \log\frac{\bar a_k(t)}{1-\nu(t) \bar a_k(t)}+
\sum_{l=0}^N\,\int_0^{a_l(t)}\frac{\rho_l(t,w)}{w- 1/\bar
a_k(t)}\,\,d\zeta,\nonumber
\end{equation}
for $k=1,...,N$. A proof, that $\tau$ diverges linearly with $t$, as
$t \to \infty$, and testing the points, $\nu$ and $0$, for stability
is the same as above with the same conclusion, that all $a_k \to 0$
for $t \to \infty$, implying that an arbitrary shape, expressed by
(9), moves toward the selected bubble (8) with $U=2$. After having
analyzed symmetric bubbles, we now move on
to bubbles that are not symmetric with respect to the channel
centerline.

\begin{figure}
\begin{center}
\includegraphics[width=0.3\textwidth]{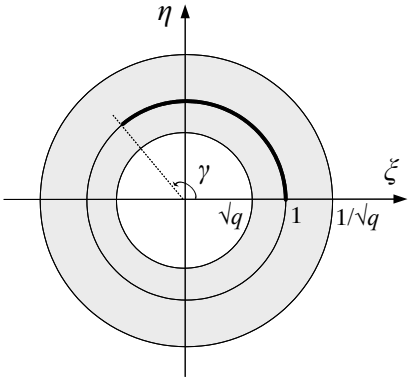}
\end{center}
\caption{Flow domain (shaded region) in the $\zeta$-plane for an asymmetric bubble; see text.}
\label{fig:asym}
\end{figure}

{\it Non-symmetric case.} The complex potential for a non-symmetric
shape requires infinity of reflected images, so we map conformally
the annulus, $0 < \sqrt q < |\zeta| <1$, in the $\zeta$-plane (see
Fig.~\ref{fig:asym}) onto the fluid domain, $D(t)$, in the
$z$-plane, so that the inner circle, $|\zeta|=\sqrt q$, is mapped
onto the interface $\Gamma(t)$. The unit circle, $|\zeta| = 1$, is
mapped onto the channel walls, $y=\pi$ and $y=0$. Under reflection
with respect to the unit circle (see Fig.2), we obtain the pre-image
of the domain, $\bar D(t)$, which is the complex conjugate of the
$D(t)$. We map
\begin{equation}
\{\sqrt q < |\zeta| < 1/\sqrt q\} \to \{ D(t) \,\cup\, \bar D(t)
\},\nonumber
\end{equation}
where the annulus $\sqrt q < |\zeta| < 1/\sqrt q$ is cut along the
part of the unit circle, $0 < \arg \, \zeta < \gamma$, so that the
inner (outer) cut side is mapped onto the upper wall (its mirror
image), where $y=\Psi=\pm\pi$, while the complimentary arc along the
unit circle, $\gamma < \arg \zeta < 2\pi$, is mapped onto the south
wall, where $y=\Psi=0$. We fix the map by sending the points
$\zeta=1$ to $x=+\infty$ and $\zeta=e^{i\gamma}$ to $x=-\infty$.

The complex potential for the non-symmetric case is
\begin{equation}
W(\zeta) = i\gamma/2 +
\log\,\frac{\Theta(e^{-i\gamma}\zeta)\,\Theta(q\zeta)}{\Theta(q
e^{-i\gamma}\zeta)\,\Theta(\zeta)},\nonumber
\end{equation}
where
\begin{equation} \Theta(\zeta) =
(1-\zeta) \prod_{m=1}^{+\infty}(1-q^{2m}\zeta)(1-q^{2m}/\zeta) =
\frac{\vartheta_4(\log(
{\sqrt{\zeta/q}}),q)}{\prod_{m=1}^{+\infty}(1-q^{2m})},\nonumber
\end{equation}
\begin{equation}
{\rm{and}}\qquad \vartheta_4(w,q) = \sum_{n=-\infty}^\infty (-1)^n
q^{n^2} \exp (2 n w) \nonumber
\end{equation}
is the Jacobi theta function \cite{Gradshtein}.
It is easy to verify
that $W(\zeta)$ satisfies the boundary conditions indicated above.

Finite-parametric solutions for the interface in this case have the
form
\begin{eqnarray}
z(t,\phi) = \tau(t) + i\gamma(t)/2 + \log\,\frac{\Theta(
e^{i(\phi - \gamma(t))})}{\Theta(e^{i\phi})} + \nonumber\\
\sum_{k=1}^N[\alpha_k \log\Theta(a_k(t)e^{-i\phi}) + \bar
\alpha_k\log\Theta({\bar a_k(t)}e^{i\phi})].
\end{eqnarray}
Here all $|a_k|<1$, and $\sum_{k=1}^N\,\alpha_k = 0$. Since $y$ is
multiple of $\pi$ when $|\zeta| = 1$, then $\tau$ is purely real.
Inserting (10) into (2) and integrating the resulting equations of
motion, we obtain $N+2$ complex constants of motion: $\beta_k = f(t,
q/\bar a_k)$ for $k=1,...,N$, $\beta_+ = f(t,q)$, and $\beta_- =
f(t,q e^{i\gamma})$, where
\begin{subequations}
\begin{align}
\beta_k  &= \tau \, + \, i\gamma/2 +
\log\,[\Theta(q\,e^{i\gamma}\bar a_k)/\Theta(q \bar a_k)] \, + \cr&
\sum_{l=1}^N \,\left[\alpha_l\,\log\,\Theta(a_l \bar a_k/q)+ \bar
\alpha_l\,\log\,\Theta(q\bar a_l/\bar a_k)\right],\\
\beta_\pm &= \tau - 2t + i\gamma/2
\pm\log\,[\Theta(q\,e^{i\gamma})/\Theta(q)] \, + \cr
&\sum_{l=1}^N\,\left[\alpha_l \log\Theta(e^{-i\gamma_\pm}\,a_l/q) +
\bar \alpha_l \log\Theta(q e^{i\gamma_\pm} \bar a_l)\right],
\end{align}
\end{subequations}
where $\gamma_+ = 1$ and $\gamma_- = \gamma$. The constants,
$\beta_+$ and $\beta_-$ are not independent, since ${\cal I}m \,
\beta_+ = {\cal I}m \, \beta_- = \gamma/2 + {\cal I}m \sum_{l=1}^N
\alpha_l \log a_l$.  The formulae (11) constitute the full dynamics
of $a_k$, $\gamma$, $q$, and $\tau$. The bubble area, $A$, while
fixed in time, is not an independent constant of motion; it is
neatly expressed through other constants as
\begin{equation} A/\pi = \beta_+ - \beta_- + 2{\cal
R}e\,\sum_{k=1}^N \bar \alpha_k \beta_k.\nonumber
\end{equation}
\begin{figure}[t]
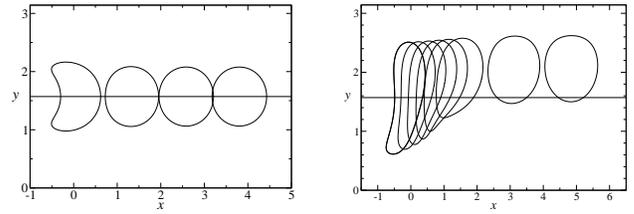

\begin{center}
\includegraphics[width=0.21\textwidth]{fig3a.eps}\qquad
\includegraphics[width=0.21\textwidth]{fig3b.eps}
\end{center}
\caption{Examples of bubble evolution: (a) symmetric solution and
(b) asymmetric solution.} \label{fig:bubble}
\end{figure}

{\it The attractor.} After eliminating $\tau$ from (11a), by
subtracting $\beta_+$  (or $\beta_-$) from $\beta_k$, we see that
the term, 2$t$, in the resulting conserved constant must be canceled
by a divergent logarithmic term. This implies that all $a_k \neq q$
must move to the point $q e^{i\gamma}$, when $t \to \infty$. The
points $q$ and $q e^{i\gamma}$ are the repeller and the attractor
respectively for the dynamical system expressed by (11). Taken at
the limit $t \to \infty$, these two points are the only fixed points
of this system.  If one of the $a_k$'s, say $a_1$, was initially at
the repeller, $a_1 = q$, then $\beta_1$ diverges, and $U =
2/(1+\alpha_1)\neq 2$, when $t \to \infty$. After relocating $a_1$
out of $q$  (by setting $a_1(0)\neq q(0)$), the asymptotic velocity
of the bubble reaches the same selected value, $U=2$, as in the
symmetric case.

For $\beta = \pi$, the solution (10) describes symmetric bubbles and
thus recovers (6), albeit in a different formulation. A symmetric
bubble evolution is shown in Fig.~\ref{fig:bubble}a, where the
asymptotic shape corresponds to the Taylor-Saffman bubble \cite{TS}
with $U=2$, as described by (8). In Fig.~\ref{fig:bubble}b we show
an asymmetric solution whose asymptotic shape coincides with the
asymmetric bubble obtained in \cite{T87,GLV2001} for $U=2$. Notice
furthermore that since all $a_k$ move toward the same point, $a_k
\to q e^{i\gamma}$, for $t \to \infty$, the sum in (11) vanishes
because $\sum_{k=1}^N \alpha_k = 0$, so $\gamma(t) \to {\cal I}m
\beta_+ = {\cal I}m \beta_- $, which is a constant of motion whose
value is set by the initial conditions.  Thus, it remains to be seen
how to centralize a bubble in our framework (if possible at all), so
that $\gamma \to \pi$ as $t\to\infty$.

For want of space, we do not present here the selection for the
asymmetric bubble with infinitely-parametric non-singular solutions,
which is the analog of (9).
It suffices to say that in this case selection of the bubble
velocity $U=2$ is also achieved in the limit $t\to \infty$.

The results presented here (and the results of \cite{1998} for a
finger, which is a singular limit of this work, when $\nu = 1$)
unambiguously indicate that the stability of the selected pattern,
with respect to the rest of the family, is built in the Laplacian
growth {\it without surface tension}.  In this context, surface
tension is just one of infinitely many perturbations (perhaps the
most relevant) which kicks the system toward the attractor, while
also regularizing high curvatures.
The selected pattern, while linearly unstable in the absence of
surface tension, is stable asymptotically: if perturbed, it
eventually recovers its original shape.

In conclusion, since the selection mechanism in both
simply-connected and doubly-connected geometries is due to the
attractor for all nonsingular solutions with zero surface tension,
we conjecture that the same holds for Laplacian growth in domains of
arbitrary connectivity.  We also think that the same selection
mechanism can be found in other free boundary problems with the same
(or similar) mathematical description
.

One of us (M.M-W) thanks MPIPKS (Dresden) and UPFE (Recife) for
hospitality
when this work was done.


\begin{thebibliography}{}
\bibitem{footnote} One should also stay away from initial data
leading to the loss of univalence of the interface. It is easy to
accommodate, and we do not discuss details here.
\bibitem{15} The rest of
formulae related to the solution (9) will be presented elsewhere.
\end{thebibliography}
\end{document}